\documentclass[a4paper,fleqn,usenatbib,twocolumn]{mnras}

\usepackage{graphicx}	
\usepackage{amsmath}	
\usepackage{amssymb}	
\usepackage{savesym}
\usepackage{natbib}

\usepackage{mathrsfs,dsfont}
\usepackage{multirow}
\usepackage{captcont,subcaption}
\usepackage{float}
\usepackage{booktabs}
\usepackage{epstopdf}
\usepackage{color}
\usepackage{hyperref}
\usepackage{verbatim}

\usepackage[none]{hyphenat}

\graphicspath{{figures/}}

\def\be{\begin{equation}}
\def\ee{\end{equation}}
\def\kms{{\rm \,km\,s^{-1}}}

\def\Gyr{{\rm \,Gyr}}

\def\Mpc{{\rm \,Mpc}}
\def\kpc{{\rm \,kpc}}

\def\msun{{\,M_\odot}}

\newcommand{\dd}{{\rm d}}

\newcommand{\MAS}{{MA-shock}}

\usepackage{xcolor}

\title[Encounters of Merger and Accretion Shocks in Clusters]{Encounters of Merger and Accretion Shocks in Galaxy Clusters and their Effects on Intracluster Medium}

\author[Congyao Zhang et al.]{
Congyao Zhang,$^{1,2}$\thanks{E-mail: cyzhang@astro.uchicago.edu}
Eugene Churazov,$^{2,3}$
Klaus Dolag,$^{4,2}$
William R. Forman,$^5$
\newauthor
Irina Zhuravleva$^1$
\\
$^1$~Department of Astronomy and Astrophysics, University of Chicago, Chicago, IL 60637, USA \\
$^2$~Max Planck Institute for Astrophysics, Karl-Schwarzschild-Str. 1, D-85741 Garching, Germany  \\
$^3$~Space Research Institute (IKI), Profsoyuznaya 84/32, Moscow 117997, Russia \\
$^4$~University Observatory Munich, Scheinerstr 1, D-81679 Munich, Germany \\
$^5$~Smithsonian Astrophysical Observatory, Harvard-Smithsonian Center for Astrophysics, 60 Garden St., Cambridge, MA 02138 \\
}

\date{Accepted XXX. Received YYY; in original form ZZZ}

\pubyear{2019}

\begin{document}
\label{firstpage}
\pagerange{\pageref{firstpage}--\pageref{lastpage}}
\maketitle

\begin{abstract}

Several types/classes of shocks naturally arise during formation and evolution of galaxy clusters. One such class is represented by  accretion shocks, associated with deceleration of infalling baryons. Such shocks, characterized by a very high Mach number, are present even in 1D models of cluster evolution. Another class is composed of ``runaway merger shocks'', which appear when a merger shock, driven by a sufficiently massive infalling subcluster, propagates away from the main-cluster center. We argue that, when the merger shock overtakes the accretion shock, a new long-living shock is formed that propagates to large distances from the main cluster (well beyond its virial radius) affecting the cold gas around the cluster. We refer to these structures as Merger-accelerated Accretion shocks (\MAS{}s) in this paper. We show examples of such \MAS{}s in 1D and 3D simulations and discuss their characteristic properties. In particular, (1) \MAS{}s shape the boundary separating the hot intracluster medium (ICM) from the unshocked gas, giving this boundary a ``flower-like'' morphology. In 3D, \MAS{}s occupy space between the dense accreting filaments. (2) Evolution of \MAS{}s highly depends on the Mach number of the runaway merger shock and the mass accretion rate parameter of the cluster. (3) \MAS{}s may lead to the misalignment of the ICM boundary and the splashback radius.

\end{abstract}

\begin{keywords}
hydrodynamics -- shock waves -- methods: numerical -- galaxies: clusters: intracluster medium
\end{keywords}


\section{Introduction} \label{sec:introduction}
The self-similar spherical collapse model provides an insightful framework for understanding the formation and evolution of galaxy clusters in the expanding Universe \citep[][]{Fillmore1984,Bertschinger1985,Adhikari2014,Shi2016b}. Characteristic sizes of both gaseous and dark matter (DM) components of galaxy clusters naturally co-exist in this model, i.e. the radius of the accretion shock $r_{\rm acc}$ \citep{Birnboim2003} and the splashback radius $r_{\rm sp}$ \citep{More2015}. They coincidentally align with each other ($r_{\rm acc}\simeq r_{\rm sp}$) if the gas adiabatic index is $\gamma=5/3$ \citep[see e.g.][]{Shi2016b}\footnote{Specifically, the alignment of the $r_{\rm acc}$ and $r_{\rm sp}$ holds when $\gamma=5/3$ only if the cluster mass accretion rate parameter ($\Gamma$) is in the range of $0.5\leq \Gamma \leq 5$ (\citealt{Shi2016b}; see the definition of $\Gamma$ in Section~\ref{sec:1d_model}).}.

However, the evolution of galaxy clusters is more complicated than those one-dimensional (1D) self-similar solutions. Two major processes tend to break the self-similarity (and also spherical symmetry) of galaxy clusters, i.e. active galactic nucleus (AGN) feedback \citep[see e.g.][for a recent review]{Werner2019} and cluster mergers \citep[e.g.][]{Sarazin2002}. The former process perturbs the gas in cluster cores (e.g. $\lesssim100\kpc$); the latter one, however, could re-distribute both gas and DM on Mpc scales \citep[e.g.][]{Ricker1998,Poole2006}. Unlike the prediction of the self-similar model, in cosmological simulations, the accretion shocks are found beyond $r_{\rm sp}$, and are sometimes even significantly farther outside  (e.g. \citealt{Lau2015,Schaal2016,Zinger2018}; see also \citealt{Walker2019} for a review). On average, in simulations $r_{\rm acc}/r_{\rm sp}$ is $\simeq1.5$ throughout the evolution of galaxy clusters (see fig.~1 in \citealt{Walker2019}). Mergers of galaxy clusters presumably play an important role in this regard.

During the merger process, the cluster splashback radius, as the outermost caustic in the DM density profile, is mainly affected through the change of the gravitational potential of the merging systems. However, due to the collisional nature of the gas, the impact of mergers on accretion shocks is more complicated. One important effect is the encounter of the merger and accretion shocks, which is able to change dramatically the shock radius \citep{Birnboim2010}. \citet{Zhang2019b} have demonstrated that the merger shocks could detach from the infalling subclusters that drive them, and propagate to large distances. They could maintain their shock strength or even get stronger when moving in the diffuse intracluster medium (ICM; say the regions between the high-density filaments), where the radial gas density profile is rather steep. In \citet{Zhang2019b}, these shocks were called ``runaway merger shocks''. On the other hand, galaxy clusters are supposed to frequently experience merger events. For example, the merger rate is a few per halo per unit redshift for the merger mass ratio $\xi \leq 30$ \citep{Fakhouri2008}. Therefore, collisions of the merger and accretion shocks could be very common.

From the theoretical point of view, the collision of two 1D shocks is a Riemann problem. Three discontinuities are subsequently formed after the shock interaction, including forward and reverse shocks/rarefactions and a contact discontinuity (CD) in between \citep{Landau1959}. More specifically, in our problem, the two colliding shocks are: a runaway merger shock with a moderate Mach number $\mathcal{M}_{\rm rs}\lesssim3$ \citep{Zhang2019b} and an accretion shock with high Mach number $\mathcal{M}_{\rm acc}\gtrsim10$ \citep[see][]{Borgani2011}, respectively. They both move radially outwards in the rest frame of the cluster. In this case, a strong forward shock is formed with  Mach number $\mathcal{M}_{\rm mas}\simeq \mathcal{M}_{\rm acc}\mathcal{M}_{\rm rs}$, moving away from the cluster center \citep{Birnboim2010}\footnote{At the same time, a reverse rarefaction wave is formed and moves towards the cluster center, which however is quickly diminished when travelling into the denser ICM region.}. In this work, it is referred to as the Merger-accelerated Accretion shock (\MAS{} hereafter).

In principle, \MAS{}s should naturally appear in all hydrodynamic cosmological simulations if their spatial resolution in the cluster outskirts is high enough to resolve the structures \citep[see e.g.][]{Miniati2000,Ryu2003,Skillman2008,Vazza2010,Vazza2011,Paul2011,Schaal2016,Zinger2016,Ha2018}. These shocks shape atmospheres of galaxy clusters, producing a ``blossom-like'' morphology (see e.g. fig.~2 in \citealt{Vazza2017}). By definition, \MAS{}s are a subset of the external shocks classified in \citet{Ryu2003}, which, however, are expected to behave quite differently from the ordinary accretion shocks. It is therefore worth studying the nature of \MAS{}s and their implications in  cluster formation. In this work, we argue that most of the time, clusters are encompassed by \MAS{}s rather than canonical accretion shocks.

This paper is organized as follows. In Section~\ref{sec:1d_model}, we illustrate formation of \MAS{}s in a 1D cluster model, and explore the evolution of the \MAS{} structures and its dependence on the cluster environment. In Section~\ref{sec:3d_simulations}, we extend our exploration to full 3-dimensional (3D) cosmological simulations, and make a direct comparison of a 3D cluster with our 1D model. In Section~\ref{sec:conclusion}, we discuss the impacts of \MAS{}s on the ICM and make conclusions.

\section{Modelling MA-Shocks} \label{sec:1d_model}

For the purposes of demonstrating the formation of \MAS{}s, we performed 1D cosmological simulations in this section. These simulations are similar to those used in \citet[][see also \citealt {Birnboim2010}]{Birnboim2003}, where both gas and DM components of the Universe are simulated in the cosmological comoving background. The 1D model is computationally fast and isolates the formation of \MAS{}s from the much more complicated merger/accretion configurations in 3D (see Section~\ref{sec:3d_simulations} for a comparison of the 1D and 3D simulation results).

Here we briefly describe the numerical method and the initial conditions used in the 1D model (see more details in Appendix~\ref{sec:appendix:1d_simulation}). Our simulations employ a hybrid N-body/hydrodynamic method \citep[see e.g.][]{Bryan1995}, where Eulerian scheme is used to solve the gas dynamics while the DM is modelled as Lagrangian shells. All our simulations presented in this section start from the redshift $z=100$. The initial gas and DM density profiles are designed so that the cluster grows in a self-similar way with a constant mass accretion parameter $\Gamma$ defined as
\be
M(t) = M_0 a(t)^\Gamma,
\label{eq:mar}
\ee
where $M_0$ is the cluster mass at present and $a(t)$ is the cosmic scale factor \citep{Fillmore1984}. We stress that, instead of applying more realistic initial conditions \citep[e.g.][]{Birnboim2003}, our choice helps to understand the relation of the \MAS{} evolution and the state of the cluster growth. Our results show that the trajectory of the \MAS{} front strongly depends on the value of $\Gamma$ (see Figs.~\ref{fig:rsh_pred}~and~\ref{fig:rsh_pred_scaled} below, and Section~\ref{sec:1d_model:discussion} for more discussions). We have also used more sophisticated initial conditions in the 1D simulations in Section~\ref{sec:3d_simulations:comparison}, and directly compare the 1D results with the 3D cosmological simulations.

To generate a \MAS{}, an additional ``merger'' shock is initiated at the cluster center at the moment $t_{\rm b}$ by suddenly increasing the gas pressure in the innermost cell by a factor of $\xi$. We vary $\xi$ to obtain ``runaway'' shocks with different Mach number $\mathcal{M}_{\rm rs}$ when they encounter the cluster accretion shock (see Table~\ref{tab:simulation_params} for a summary of the main parameters used in our simulations). This method has been used and proved to be robust in \citet{Zhang2019b} when they studied the propagation of runaway merger shocks in cluster outskirts.

\begin{table}
\centering
\begin{minipage}{0.45\textwidth}
\centering
\caption{Parameters of 1D cosmological simulations.}
\label{tab:simulation_params}
\begin{tabular}{*{5}{c}}
  \hline\hline
  IDs &
  $\Gamma$\footnote{The cluster mass accretion rate parameter used to set-up the initial gas/DM density profiles.} &
  $t_{\rm b}$ (Gyr)\footnote{The time a blast wave is initiated at the cluster center. No blast wave is included in the runs S1 and S3.} &
  $t_{\rm mas}$ (Gyr)\footnote{The time the \MAS{} is formed. Its uncertainty is $0.05\Gyr$, determined by the time interval between two successive snapshots. } &
  $\mathcal{M}_{\rm rs}$\footnote{The Mach number of the ``runaway'' shock at the moment just before it encounters the accretion shock.} \\\hline
  S1      & $1$ & -     & -   & -   \\
  S1T2M15 & $1$ & $2.5$ & 3.0 & 1.5 \\
  S1T2M20 & $1$ & $2.5$ & 2.8 & 2.0 \\
  S1T2M23 & $1$ & $2.5$ & 2.8 & 2.3 \\
  S1T6M23 & $1$ & $6.5$ & 6.7 & 2.3 \\
  S3      & $3$ & -     & -   & -   \\
  S3T2M20 & $3$ & $2.5$ & 2.7 & 2.0 \\
  S3T2M25 & $3$ & $2.5$ & 2.7 & 2.5 \\
  S3T6M20 & $3$ & $6.5$ & 7.0 & 2.0 \\
\hline\hline
\vspace*{-5mm}
\end{tabular}
\end{minipage}
\end{table}

\subsection{Formation of \MAS{}s and their structures} \label{sec:1d_model:formation}

\begin{figure}
\centering
\includegraphics[width=\linewidth]{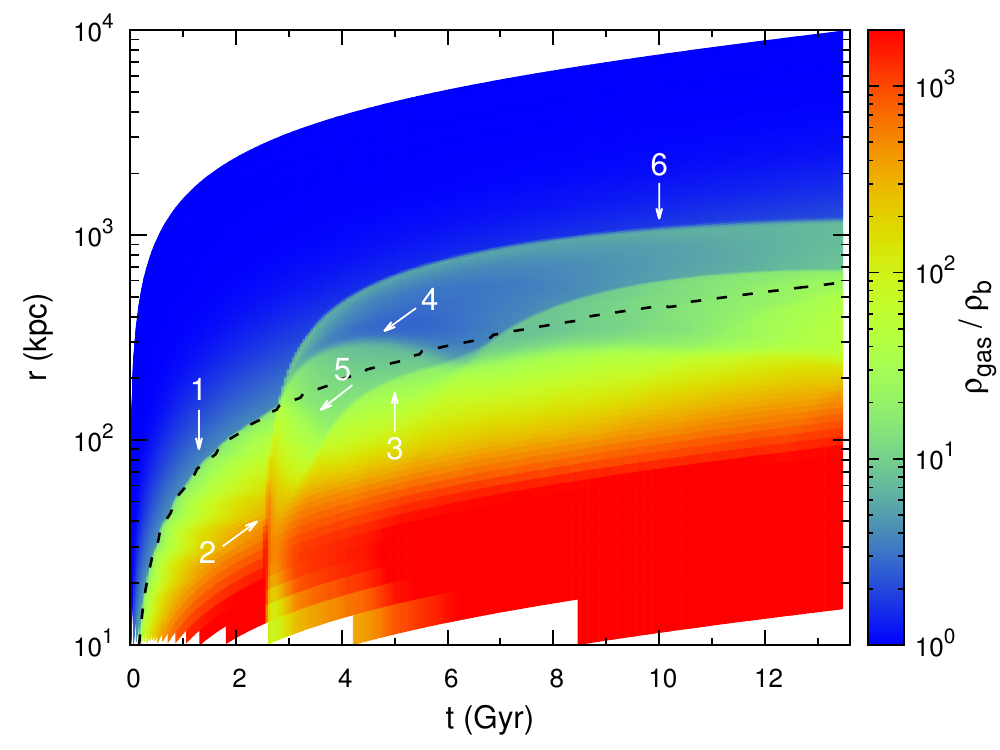}
\caption{Evolution of gas density profile in the simulation S1T2M23, where the gas density $\rho_{\rm gas}$ is scaled by the cosmic mean density of baryons $\rho_{\rm b}$. The ``runaway'' shock is initiated at the cluster center at $t=2.5\Gyr$ and encounters the accretion shock at $t_{\rm mas}\simeq2.7\Gyr$. A rarefaction, CD, and \MAS{} are subsequently formed after the shock collision. The gaseous structures, including: 1.~original accretion shock, 2.~front of the runaway shock, 3.~rear of the runaway shock, 4.~CD, 5.~rarefaction, 6.~\MAS{}, are marked in the figure by their corresponding numbers. As a comparison, the evolution of the accretion shock radius in the run S1 is shown as the black dashed line. This figure shows several new structures form after the collision of the merger and accretion shocks. The boundary of the shock-heated cluster atmosphere is driven much farther outwards by this collision (see Section~\ref{sec:1d_model:formation}). }
\label{fig:1d_map}
\end{figure}

Fig.~\ref{fig:1d_map} shows the evolution of the gas density profile in the simulation S1T2M23. The initial accretion shock forms at the very beginning of the simulations once the collapsing gas decouples from the Hubble flow. Since there is no radiative cooling involved in the simulations, the stable gas atmosphere could exist even when the cluster mass is small (cf. \citealt{Rees1977,Birnboim2003}). A secondary shock is artificially initiated at the cluster center at $t=2.5\Gyr$ to mimic a merger shock, which propagates with a high speed in the ICM and rapidly catches up with the accretion shock. As expected, a rarefaction, CD, and \MAS{} are formed after the shock collision (marked in the figure; see also \citealt{Birnboim2010}). Both CD and \MAS{} are subsequently decelerated by the inflowing unshocked gas. Their trajectories, however, depend on the environment of the cluster (i.e. mass accretion rate parameter $\Gamma$) and the strength of the \MAS{} (cf. Fig.~\ref{fig:1d_map_appendix}; see Section~\ref{sec:1d_model:evolution} for more detailed discussions on these dependence). When the $\Gamma$ is moderate (like the case $\Gamma=1$ shown in Fig.~\ref{fig:1d_map}), a long time is needed for the \MAS{} to re-fall back (even possibly longer than the Hubble time). It is interesting to note that the runaway shock in our model develops an N-shaped wave profile due to its blast-wave nature and the spherical symmetry of the system \citep{Dumond1946}\footnote{A similar N-wave form of the runaway merger shock is also seen in our idealized simulations of the mergers between two galaxy clusters (see the right-most panel in fig.~1 in \citealt{Zhang2019b}).}. The front of the N-wave encounters the accretion shock at $t\simeq2.7\Gyr$. The rear part of the N-wave, however, firstly meets the re-infalling CD at $t\simeq6\Gyr$. Its velocity increases significantly after crossing the CD from the cold side to the hot side.

\begin{figure}
\centering
\includegraphics[width=\linewidth]{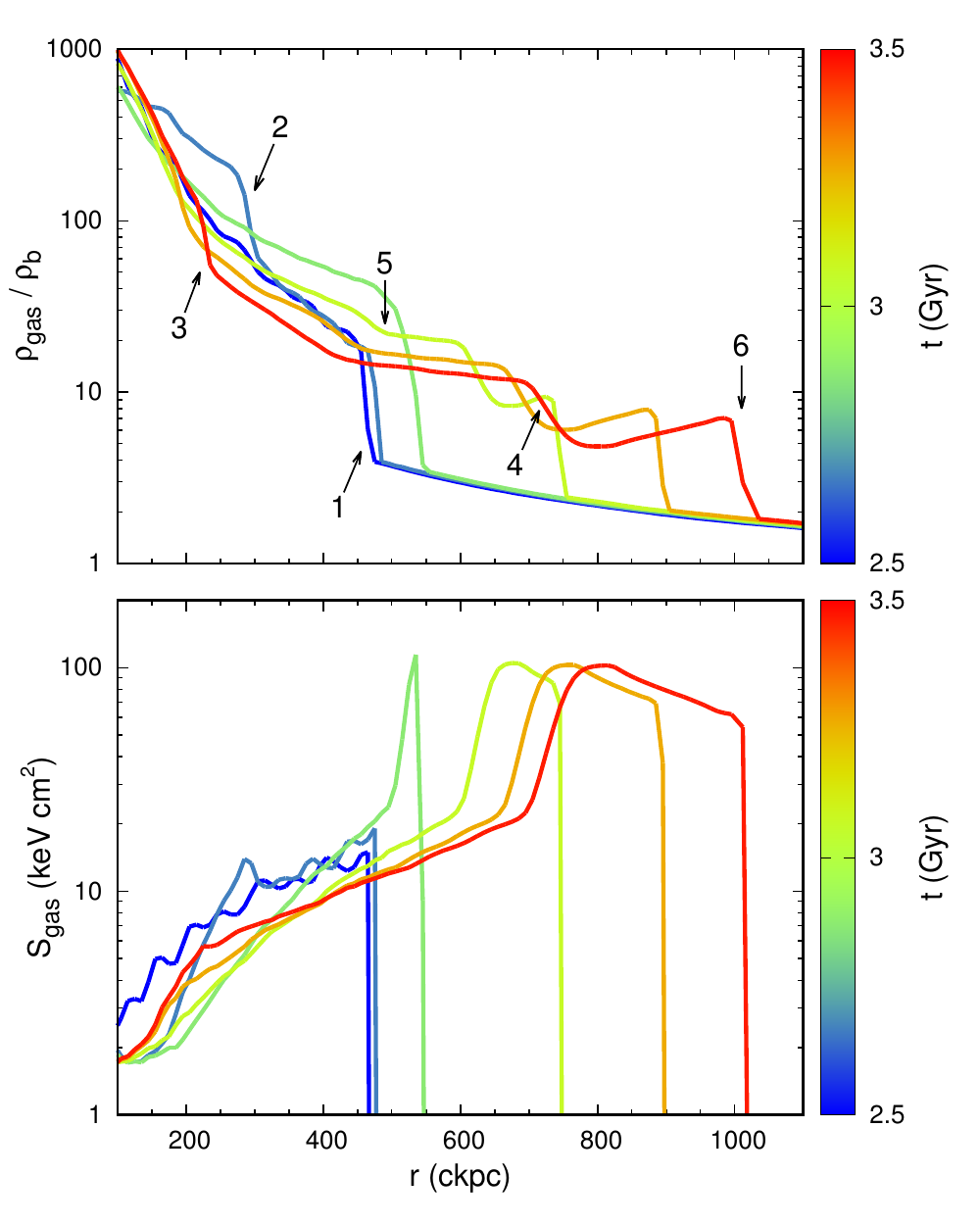}
\caption{Profiles of the gas density (top panel) and entropy (bottom panel) in the simulation S1T2M23. Note that the $x$-axis is in comoving coordinates (different from that shown in the bottom panel of Fig.~\ref{fig:1d_map_appendix}).  The line colors encode the cosmic time. The numbers marking the gaseous structures correspond to those shown in Fig.~\ref{fig:1d_map}. This figure presents a clear view of the formation of \MAS{}. The high-entropy shell between the CD and \MAS{} is unambiguous evidence of the past shock collision (see Section~\ref{sec:1d_model:formation}). }
\label{fig:1d_profiles}
\end{figure}

To illustrate the structures associated with the \MAS{} more clearly, we zoom in on gas profiles where the \MAS{} forms, shown in Fig.~\ref{fig:1d_profiles}. We can clearly see the aforementioned rarefaction, CD, and \MAS{} in the gas density profiles. The CD separates the low- and high-entropy gas on its two sides. On the left (radii smaller than CD) side the gas is successively compressed and heated by the runaway and accretion shocks, while on the right side (outside CD) the gas passes only through the \MAS{}, which has velocity comparable to the merger shock velocity. As a result, the gas density is higher but the temperature is lower on the left side of CD. Therefore, a high-entropy gas shell is formed between the CD and \MAS{}, and this shell is a robust signature of the past shock collision.

It turns out that, within this high-entropy shell, the entropy is decreasing with radius. The entropy here is defined as $S_{\rm gas}\equiv T_{\rm gas}/\rho_{\rm gas}^{\gamma-1}$ ($T_{\rm gas}$ and $\rho_{\rm gas}$ are gas temperature and density, respectively). The gas temperature behind the shock is proportional to $u_{\rm mas}^2$, where $u_{\rm mas}$ is the \MAS{} velocity, which is a decreasing function of time/radius for a propagating spherical shock (see Fig.~\ref{fig:vsh_evo}, and more discussions in Section~\ref{sec:1d_model:evolution}). The upstream cold gas density also decreases with the radius, but rather slowly (approximately $\rho_{\rm gas}\propto r^{-1}$ before entering the ICM; see also fig.~1 in \citealt{Shi2016b}). The net result of these competing effects is that the gas entropy profile between the CD and \MAS{} is a decreasing function of radius (see bottom panels in Figs.~\ref{fig:1d_profiles}~and~\ref{fig:1d_map_appendix}).

Given the radially decreasing entropy profile, the high-entropy shell is convectively unstable (even though this instability is not captured in the 1D simulations). We note in passing that the temperature is also decreasing with radius and, therefore, the shell is unstable even when the magnetothermal-instability criterion \citep{Balbus2000} is used instead of the Schwarzschild one mentioned above. With time, convective motions should kick in and the shell would re-arrange itself to restore the non-decreasing entropy profile. The instability growth rate can be related to the characteristic Keplerian frequency ($\Omega_{\rm K}$) as $\sim\Omega_{\rm K} \sqrt{\gamma^{-1}\dd\ln S_{\rm gas}/\dd \ln r}$, implying that the life-time of \MAS{}s is likely shorter than that of the shell. Our estimates show that, given the slope of the entropy profile and its radial extent, the characteristic amplitude of the induced gas motions can be up to $\sim 0.5c_{\rm s}$, where $c_{\rm s}$ is the ICM sound speed. This implies that the resulting convection might make an important contribution to the non-thermal pressure in the cluster outskirts \citep[see also][]{Shi2014,Shi2015}.

\subsection{Evolution of \MAS{}s and their fate} \label{sec:1d_model:evolution}

The evolution of \MAS{}s shows a very similar behavior to that of a strong blast wave \citep[see e.g.][]{Ostriker1988}. The accreted cold gas on the upstream side strongly decelerates the shocks.
The velocity of the \MAS{} front in the cluster frame can be represented as a sum of the shock velocity in the rest frame of the upstream gas $u_{\rm mas}(t)$ and the infalling velocity of the upstream gas $u_{\rm gas}(r,\,t)$,
i.e.
\be
\frac{\dd r_{\rm mas}}{\dd t} = u_{\rm mas}(t) + u_{\rm gas}(r_{\rm mas},\, t),
\label{eq:traj_rmas}
\ee
where $r_{\rm mas}$ is the location of the \MAS{}s front. Generally, there are three possible fates for a \MAS{}, i.e. (1) it survives until $z=0$ (like the case shown in Fig.~\ref{fig:1d_map}); (2) it recedes and is eventually replaced with an ordinary accretion shock (like the case shown in Fig.~\ref{fig:1d_map_appendix}); and (3) it is overrun by a new runaway merger shock and is accelerated again. To keep the problem simple, we merely address the first two possibilities in this section, and will discuss the last one in Section~\ref{sec:3d_simulations}.

\begin{figure}
\centering
\includegraphics[width=\linewidth]{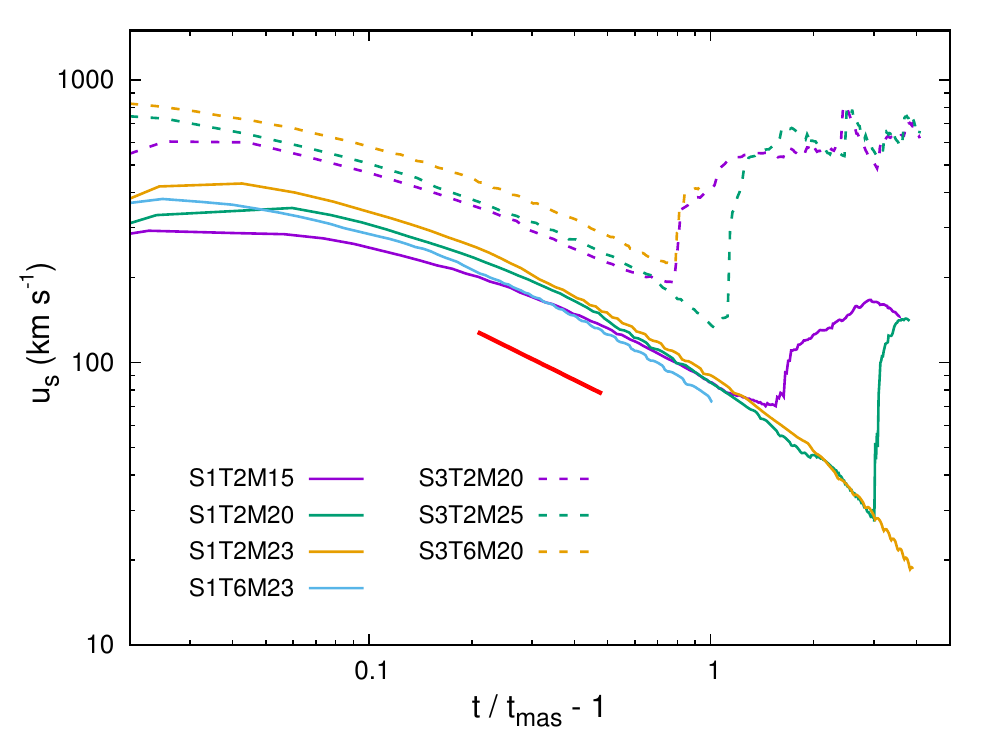}
\caption{Evolution of shock velocities of the radially outermost shocks in the simulations. The results are shown as the solid and dashed lines for the simulations with $\Gamma=1$ and $3$, respectively. The additional thick red line shows $\propto (\Delta t/t_{\rm mas})^{-3/5}$ for a comparison. This figure shows that the evolution of \MAS{}s is close to a Sedov-Taylor solution. The shape of the scaled velocity curves only mildly depends on $\Gamma$, $t_{\rm mas}$, and $\mathcal{M}_{\rm rs}$ (see Section~\ref{sec:1d_model:evolution}).}

\label{fig:vsh_evo}
\end{figure}

Fig.~\ref{fig:vsh_evo} shows the evolution of the shock velocities of the radially outermost shocks in our simulations. The \MAS{}s behave differently from that of the genuine accretion shocks.
The latter one follows $u_{\rm acc}\propto t^{\delta}$, where $\delta=(2\Gamma/3-1)/3$ \citep{Fillmore1984,Shi2016b}. The evolution of \MAS{}s, however, is close to a Sedov-Taylor solution, i.e. $u_{\rm s}\propto (\Delta t/t_{\rm mas})^{-3/5}$ for not too small $\Delta t/t_{\rm mas}$, where $\Delta t=t-t_{\rm mas}$ \citep[see][]{Sedov1959}. This is not surprising, since the \MAS{}s are driven from inside, and have negligible upstream gas pressure. Note that the \MAS{}s, when viewed as a function of $\Delta t=t-t_{\rm mas}$, go through a transitional stage before approaching the Sedov-Taylor form (i.e. $\Delta t/t_{\rm mas}\lesssim0.1$ in our simulations), when the shock velocity is changing slowly. This is because, for small $\Delta t/t_{\rm mas}$, the \MAS{}s, which do not start from the cluster center, behave more like plane shock waves rather than spherical ones in their very early phase. In addition, the evolution of \MAS{}s is also affected by (1) the gravity, (2) the Hubble expansion, and (3) non-uniform density distribution of the upstream gas, which cause slight deviations of the shock velocity curves from a simple power law. Eventually, some \MAS{}s become the accretion shocks again (i.e. the shock velocity curves jump to high values). From the simulations, we find that, the deceleration of the \MAS{}s (say the evolution of the shock velocity with time) only mildly depends on its formation time $t_{\rm mas}$, runaway shock Mach number $\mathcal{M}_{\rm rs}$, and the mass accretion rate parameter $\Gamma$. These allow us to model $u_{\rm mas}(t)$ by a very simple form, i.e.
\be
u_{\rm mas}(t) =
\begin{cases}
    u_{\rm mas}(t_{\rm mas}) & 0\leq \frac{\Delta t}{t_{\rm mas}}\leq \kappa \\
    u_{\rm mas}(t_{\rm mas})\left(\frac{\Delta t}{\kappa t_{\rm mas}}\right)^{\beta} & \frac{\Delta t}{t_{\rm mas}}>\kappa \\
\end{cases},
\ee
where $\beta=-0.7$ (slightly steeper than $-3/5$) and $\kappa$ is a free parameter to take the shock transitional stage into account. With this approximation, the integration of Eq.~(\ref{eq:traj_rmas}) becomes straightforward. The self-similar solution of the spherical collapse model is applied to estimate the gas velocity $u_{\rm gas}(r,\,t)$ \citep{Bertschinger1985,Shi2016b}.

\begin{figure}
\centering
\includegraphics[width=\linewidth]{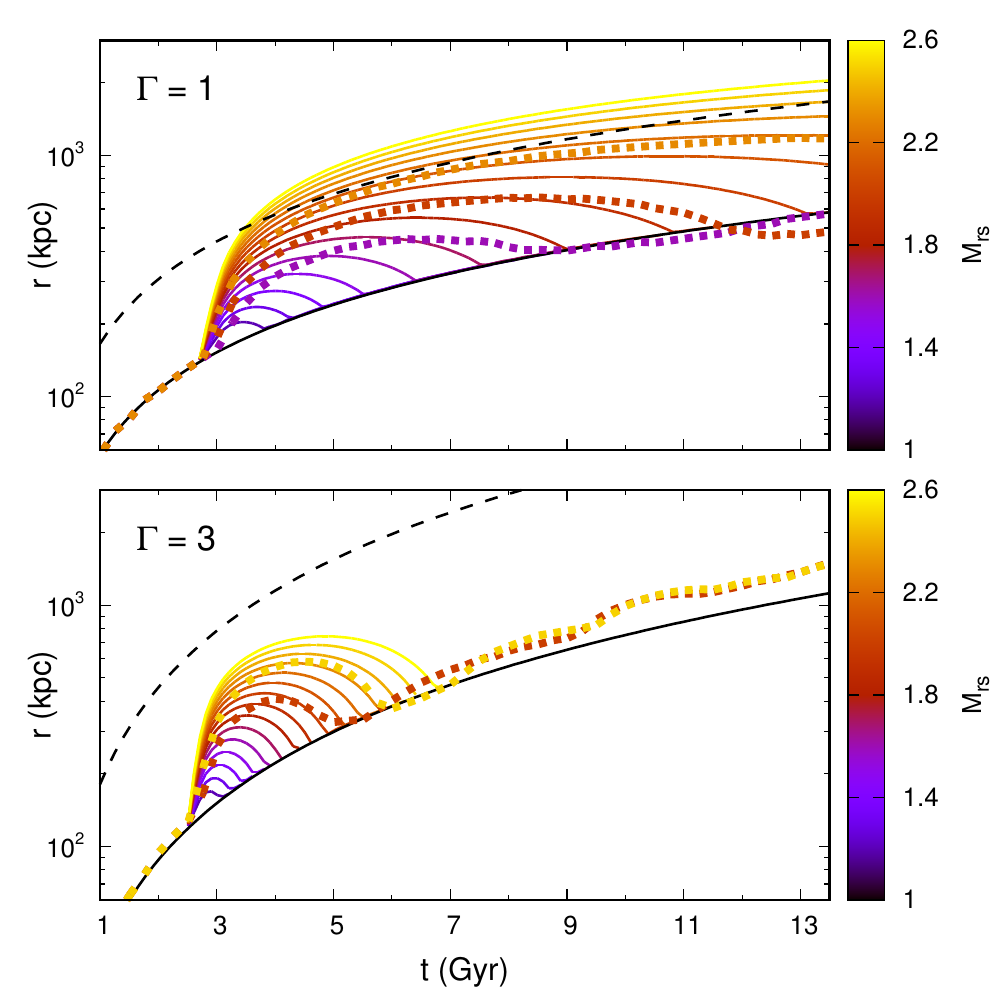}
\caption{Trajectories of the radially outermost shocks in our 1D simulations (dashed lines) and the analytical models (color solid lines). The line color encodes the Mach number of the runaway shocks $\mathcal{M}_{\rm rs}$ right before they encounter the accretion shocks. As a comparison, the black solid and dashed lines show the evolution of the accretion shock radius and the turnaround radius based on the self-similar model. This figure shows a significant dependence of the \MAS{} evolution on the value of $\Gamma$ (see Section~\ref{sec:1d_model:evolution}).  }

\label{fig:rsh_pred}
\end{figure}

Fig.~\ref{fig:rsh_pred} shows two groups of solutions of Eq.~(\ref{eq:traj_rmas}) (see Table~\ref{tab:model_params} for the parameters used in the calculations). The line color encodes the Mach number of the runaway shocks $\mathcal{M}_{\rm rs}$ just before they encounter the accretion shocks. As a reference, the evolution of the accretion shock radius $r_{\rm acc}(t)$ and the turnaround radius $r_{\rm ta}(t)$ in the self-similar solutions are shown as the black solid and dashed lines. One can see the modelled shock-front trajectories (color solid lines) show a good match with the simulations (dotted lines)\footnote{Though not shown in figures, the analytical model matches the results of the simulation runs S1T6M23 and S3T6M20 as well.}.
The \MAS{} could survive for a longer time and reach a larger radius when $\Gamma$ is smaller and/or the shock Mach number is higher. All the \MAS{}s tend to move back to the radius $r_{\rm acc}(t)$ in Fig.~\ref{fig:rsh_pred}, even though in some cases the required timescale is longer than the Hubble time. In principle, if strong enough, a \MAS{} is able to escape from the halo potential well (e.g. go beyond a few turnaround radii of the halo) and move into a nearly uniform but cosmologically expanding medium. In this case, the shock is described by the cosmological self-similar solution of blast waves found by \citet{Bertschinger1983}, where the shock radius evolves as $r_{\rm mas}\propto \Delta t^{4/5}$. However, because the Mach number of merger shocks is usually $\lesssim3$, such a situation must be very rare.

\begin{table}
\centering
\begin{minipage}{0.45\textwidth}
\centering
\caption{Parameters used in Eq.~(\ref{eq:traj_rmas}) for calculation of \MAS{} trajectories (see Section~\ref{sec:1d_model:evolution}).}
\label{tab:model_params}
\begin{tabular}{*{6}{c}}
  \hline\hline
  IDs &
  $\Gamma$\footnote{The mass accretion rate.} &
  $t_{\rm mas}$ (Gyr)\footnote{The moment \MAS{} forms.} &
  $r_{\rm mas}(t_{\rm mas})$ (kpc)\footnote{The radius where \MAS{} forms at $t=t_{\rm mas}$.} &
  $\kappa$\footnote{\MAS{} starts to behave in a self-similar way at $t=(1+\kappa) t_{\rm mas}$.} & \\\hline
  G1 & $1$ & $2.7$ & 140 & $0.12$ \\
  G2 & $3$ & $2.5$ & 120 & $0.08$ \\
\hline\hline
\vspace{-5mm}
\end{tabular}
\end{minipage}
\end{table}

\begin{figure}
\centering
\includegraphics[width=\linewidth]{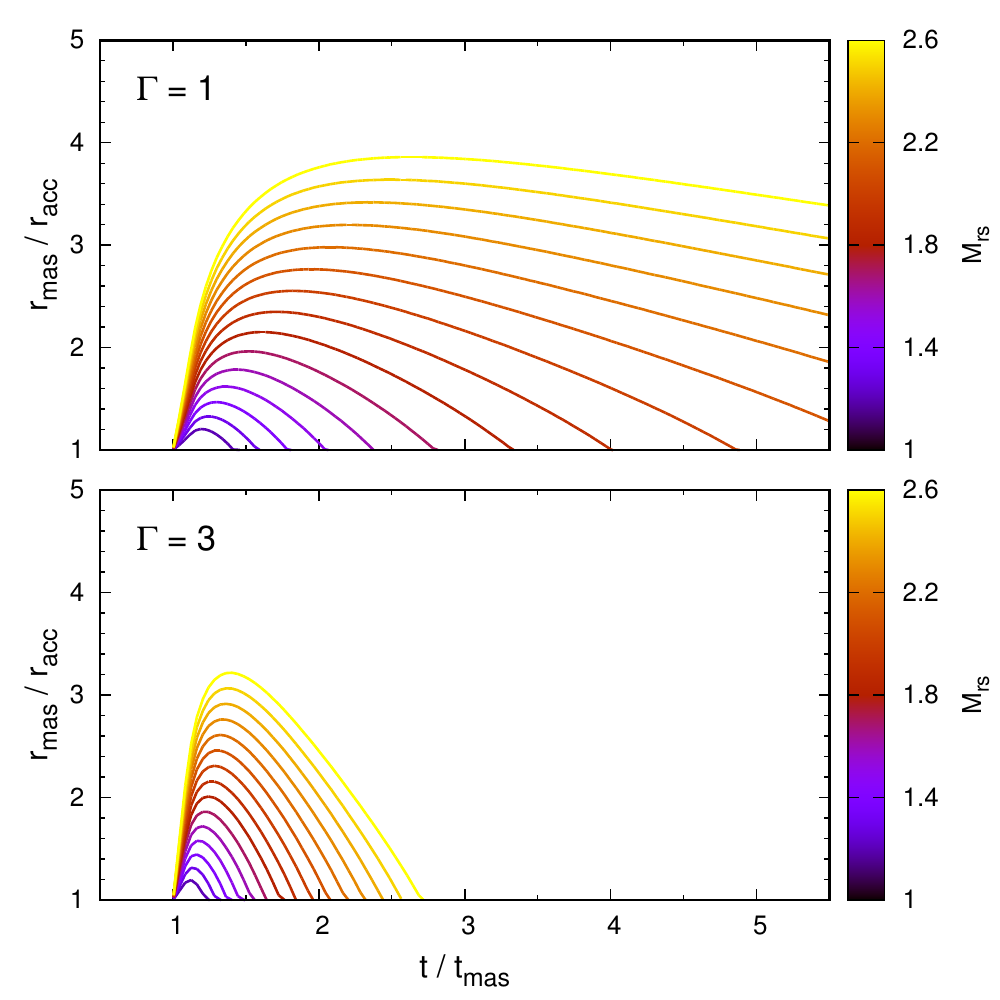}
\caption{The same data shown as the color solid lines in Fig.~\ref{fig:rsh_pred}, but the radius- and time-axis are scaled by the accretion shock radius $r_{\rm acc}$ in the self-similar model and the cosmic time $t_{\rm mas}$ when the \MAS{} forms, respectively. This figure shows that the \MAS{} could reach a larger cluster radius and survive for a longer time if $\Gamma$ is smaller and/or the Mach number of the runaway shock is higher (see Section~\ref{sec:1d_model:evolution}).}
\label{fig:rsh_pred_scaled}
\end{figure}

Fig.~\ref{fig:rsh_pred_scaled} shows the same results as those in Fig.~\ref{fig:rsh_pred} but with $r_{\rm mas}$ and $t$ scaled by the accretion shock radius $r_{\rm acc}$ in the self-similar model and the time $t_{\rm mas}$ when forming the \MAS{}s\footnote{Here we only show the results when $t_{\rm mas}\simeq3\Gyr$ (see Table~\ref{tab:model_params}). But we have compared the models with $t_{\rm mas}\simeq3$ and $7\Gyr$, and found the curves show only weak dependence on $t_{\rm mas}$.}. This figure shows that, if $\Gamma=1$, even a moderate runaway merger shock (e.g. $\mathcal{M}_{\rm rs}\gtrsim1.5$) could easily expand the size of atmospheres of galaxy clusters by a factor of $\gtrsim2$, and the expansion could last for a few $t_{\rm mas}$. However, when $\Gamma$ gets larger, it is harder for the \MAS{}s to survive for longer time and reach larger cluster radii. The dependence of the upstream gas velocity profile $u_{\rm gas}$ on $\Gamma$ plays a key role in this result (see Eq.~\ref{eq:traj_rmas}). For example, when $\Gamma=3$, the \MAS{}s produced by the runaway shocks with $\mathcal{M}_{\rm rs}>2$ could only survive for a time period $\simeq t_{\rm mas}$. Nevertheless, those \MAS{}s in galaxy clusters with high $\Gamma$ would still make a significant impact on the ICM at low redshift (e.g. $z\lesssim1$).

\subsection{Key parameters $\Gamma$ and $\mathcal{M}_{\rm rs}$ shaping \MAS{}s} \label{sec:1d_model:discussion}

The two parameters -- the mass accretion rate $\Gamma$ and the Mach number of a runaway shock $\mathcal{M}_{\rm rs}$ play crucial roles in shaping \MAS{}s in our simulations, and are also tightly related with the cluster environment and growth history. In this section, we discuss them in detail.

The mass accretion rate that quantifies the global growth history of galaxy clusters, has been extensively explored in analytical and numerical studies \citep[e.g.][]{Zhao2009,Adhikari2014,Lau2015}. The infalling substructures and filaments contribute a large portion of the accreting mass (see the high peaks in the curves shown in the bottom panel of Fig.~\ref{fig:rsh_cluster}). However, we have to emphasize that the accretion rate defined there has a different meaning from the one used in our 1D model. Instead of characterizing the mass growth of a whole galaxy cluster, our parameter $\Gamma$ is used to set-up the matter distribution in the model. \citet{Zhang2019b} has argued that  prominent runaway merger shocks may only appear in the non-filamentary regions, and so do the \MAS{}s. In this sense, the constant $\Gamma$ applied in our 1D simulations in Section~\ref{sec:1d_model} is supposed to represent the non-filamentary environment of the cluster rather than the total mass accretion rate of the cluster integrated over all directions.

Fig.~\ref{fig:rsh_pred_scaled} implies that runaway merger shocks with  moderate Mach number (e.g. $\mathcal{M}_{\rm rs}\gtrsim1.5$) are generally required to generate relatively long-lived and significant \MAS{}s. However, not all cluster mergers are powerful enough to drive such shocks. For example, one necessary condition is that the infalling subcluster should be sufficiently massive to keep its gas atmosphere after the primary pericentric passage. In this regard, the merger mass ratio $\xi$ is a key factor. We have explored the cluster merger process for a wide range of merger parameters in our previous works \citep[see e.g.][]{Zhang2014,Zhang2016}\footnote{In those works, we simulated mergers between two idealized galaxy clusters, where each of the merging clusters contains spherical gas and DM halos. The initial gas and DM density profiles both follow $r^{-3}$ in the cluster outskirts (see eqs.~1-4 in \citealt{Zhang2014}). A large merger-parameter space (including the cluster mass, mass ratio, initial relative velocity, and impact parameter) has been explored by those simulations.},
which provided some intuitions on this question. For example, \citet{Zhang2019a} and \citet{Lyskova2019} presented analysis of two merging systems with $\xi=10$ and zero and large impact parameters respectively. In both cases, the merger shocks could reach the cluster outskirts with Mach number larger than 2. Our merger sample with $\xi=60$ and zero impact parameter, however, shows that only a weak runaway merger shock (Mach number $\sim1.4$) is formed during the merger process. Overall, these results imply that $\xi\lesssim50$ is a reasonable mass-ratio range for cluster mergers to power significant runaway merger shocks. However, we emphasize that, besides the mass ratio, many other factors (like other merger parameters, initial cluster gas/DM profiles) could also affect the conclusion. To fully characterize this problem, more detailed studies need to be carried out in the future. Nevertheless, we suggest that mergers with mass ratio $\lesssim$ a few 10s would be able to drive runaway merger shocks with $\mathcal{M}_{\rm rs}\gtrsim1.5$, and further excite \MAS{}s.

\section{MA-Shocks in 3D Cosmological Simulations} \label{sec:3d_simulations}

Even though \MAS{} formation has been well captured in our 1D models, we extend our study of \MAS{}s into the 3D simulations in this section for the following reasons. (1) Galaxy clusters gradually become asymmetric at large radii. Giant filaments penetrate into the ICM along some directions. (2) Only a single merger event is considered in our 1D models. But in reality, it is very common for galaxy clusters to experience multiple merger events in a short-time period.

We analyzed a galaxy cluster from the COMPASS\footnote{www.magneticum.org/complements.html\#Compass} zoom-in simulations. This simulation is a very high resolution re-simulation of the {\it D.17} region as introduced in \citet{Bonafede2011}. The cluster's virial radius and virial mass are $R_{\rm 200m}=4.2\Mpc$ and $M_{\rm 200m}=2.2\times10^{15}\msun$\footnote{$R_{\rm 200m}$ is referred to as the virial radius of the cluster in this work, which encloses an average matter density $200$ times higher than the mean matter density of the Universe. $M_{\rm 200m}$ is the cluster virial mass within $R_{\rm 200m}$.}, respectively. It is simulated using P-Gadget3, a modernized version of P-Gadget2 \citep{Springel2005}, that implements updated smoothed particle hydrodynamic (SPH) formulations regarding the treatment of viscosity and the use of kernels \citep{Dolag2005,Beck2016}, allowing a better treatment of turbulence within the ICM. It also includes a formulation of isotropic, thermal conduction at 1/20th of the classical Spitzer value \citep{Spitzer1962}. 
The particle mass for DM and gas is $4.7\times10^6\msun$ and $8.9\times10^5\msun$, respectively, and the softening for both, DM and gas particles is set to $0.69\kpc$. The cluster at redshift $z=0$ is therefore resolved with $5.6\times10^8$ particles within the virial radius and hosts almost $10^5$ identified substructures, making it to one of the most resolved, cosmological, hydrodynamical simulations of massive galaxy clusters.

\subsection{\MAS{}s in a 3D cluster} \label{sec:3d_simulations:3d}

\begin{figure}
\centering
\includegraphics[width=\linewidth]{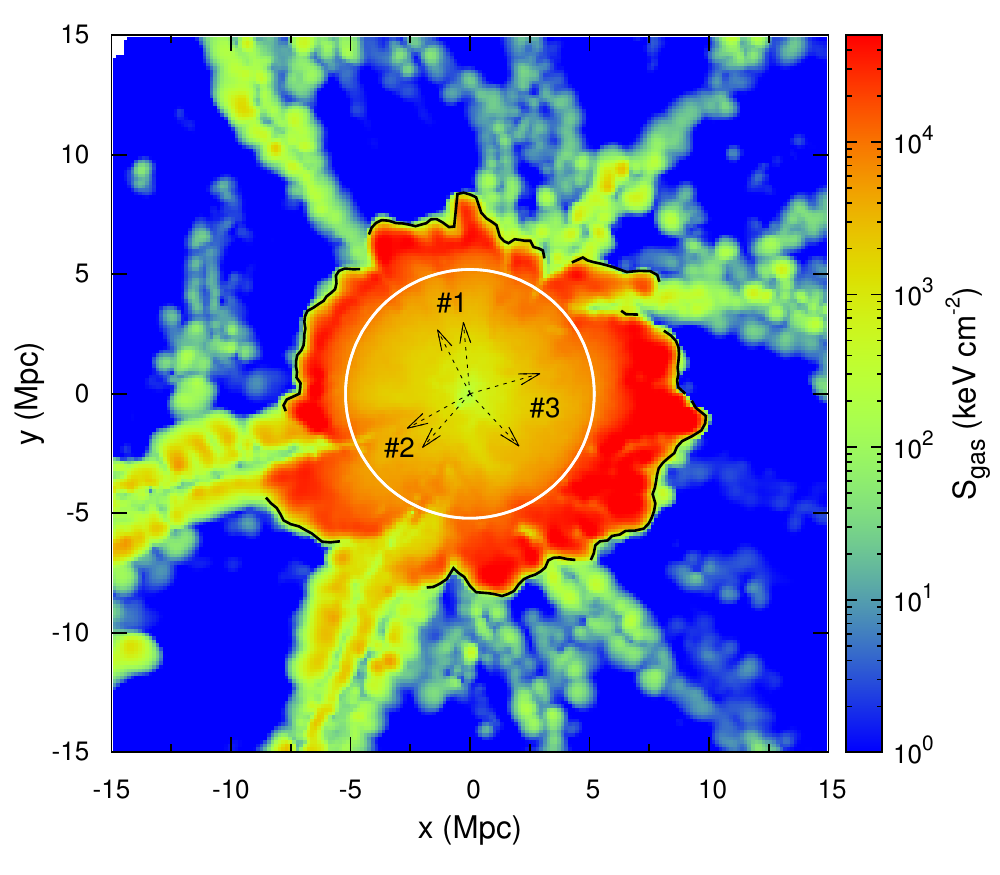}
\caption{Gas-entropy slice of a galaxy cluster from the 3D cosmological simulation at redshift $z=0$. The black solid lines highlight the \MAS{}s (or accretion shocks) identified in this map. Three different sectors are selected (marked by dashed arrows), where evolution of the shock radii are traced and shown in Fig.~\ref{fig:rsh_cluster}. The white circle shows the shock radius formed in our special 1D simulation (see Section~\ref{sec:3d_simulations:comparison}), which is used to directly compare to the 3D cluster shown here. This figure shows that the ICM in this cluster is mostly covered by \MAS{}s along the non-filamentary directions (see Section~\ref{sec:3d_simulations}). }
\label{fig:3d_map}
\end{figure}

Fig.~\ref{fig:3d_map} shows a gas-entropy slice of the galaxy cluster at the redshift $z=0$.  The black solid lines mark the positions of the \MAS{}s (or accretion shocks) identified in this map. We exclude the shocks in the direction of filaments, where the shock structures are complicated and are sometimes even pushed inside the virial radius by the strong inflowing gas streams \citep{Zinger2016}. In this figure, we can clearly see a high-entropy shell lying on the downstream side of the shock predicted by our 1D models, which implies that the cluster gas atmosphere is mostly covered by the \MAS{}s but not the genuine accretion shocks along the non-filamentary directions.

\begin{figure}
\centering
\includegraphics[width=\linewidth]{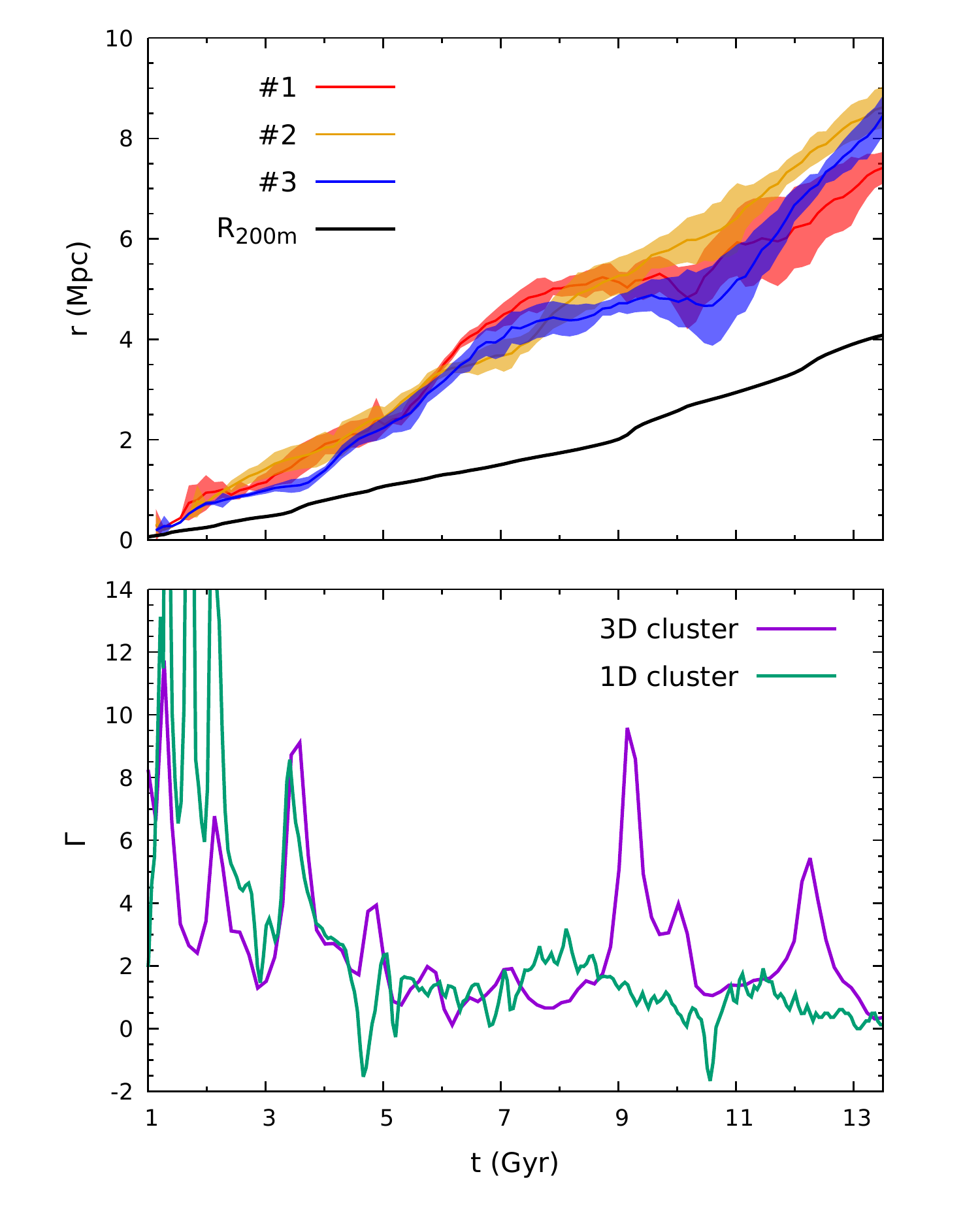}
\caption{\emph{Top panel:} Evolution of the shock radii averaged within the three sectors shown and labelled in Fig.~\ref{fig:3d_map}. The shaded band shows the $1\sigma$ scatter. The black line shows the evolution of the cluster virial radius $R_{\rm 200m}$. \emph{Bottom panel:} A comparison of the global mass accretion rate $\Gamma$ of the 3D cluster (purple line) and the 1D model (green line; see Section~\ref{sec:3d_simulations:comparison}). The 3D cluster experiences three-major merger events after $t>2\Gyr$. This figure shows that the radially outermost shocks of the cluster reside much beyond the virial radius. The increases of these shock radii associate with the merger events but with a $\sim1-2\Gyr$ time delay (see Section~\ref{sec:3d_simulations:3d}).  }
\label{fig:rsh_cluster}
\end{figure}

We further trace the evolution of the averaged radii of the \MAS{}s throughout the simulation within three sectors illustrated in Fig.~\ref{fig:3d_map}. These sectors are selected as they are not affected by the large-scale filaments. The results are shown in the top panel of Fig.~\ref{fig:rsh_cluster}. One can see that the shock radii are located much beyond the cluster virial radius (black solid line) along all three directions. After $t\gtrsim2\Gyr$ (or $z\lesssim3$), the cluster experiences three-major merger events (merger mass ratio $\xi\leq5$; see three major peaks in $\Gamma$ shown in the bottom panel of Fig.~\ref{fig:rsh_cluster}). The merger mass ratios are $\simeq1,\ 3,\ 5$, respectively. This merger rate is close to the averaged cluster major merger rate found in \citet[][see their figs.~7 and 8]{Fakhouri2008}. Thus, our cluster is representative to illustrate the effect of \MAS{}s on the ICM for the purpose of this work. During these three mergers, the cluster virial radius shows rapid increases. At the same time, the variations of the \MAS{} radii show a temporal correlation with that of the virial radius (also with the merger events). Two rapid increases of the shock radii start at $t\simeq5$ and $11\Gyr$, which are approximately $\sim1-2\Gyr$ later than the corresponding mergers. This is generally consistent with the timescale of the merger shock crossing the cluster radius, i.e. $r_{\rm mas}/(\mathcal{M}_{\rm rs}c_{\rm s})$. This correlation indicates that the cluster mergers play a dominant role in pushing the shock outwards.

\subsection{A direct comparison of 1D and 3D clusters} \label{sec:3d_simulations:comparison}

In this section, we make a direct comparison of the 3D cluster presented in Section~\ref{sec:3d_simulations:3d} with our 1D model to further clarify the effect of the mergers on the formation of \MAS{}s. Ideally, we need a 1D model that reproduces the growth history of the 3D cluster but suppresses the effect of the merger process on the ICM (i.e. only radial accretion is included). For this purpose, we performed a 1D simulation with a specially designed initial condition. The gas (and/or DM) density, pressure, and velocity radial profiles of the 3D cluster at redshift $z=160$\footnote{Specifically, there is no halo structure at this redshift. We select the position of the peak of the strongest perturbation in the snapshot as the origin of the radial profiles. The main progenitor of the present cluster will form here at a later time.} are directly used as the corresponding initial gas/DM profiles in the 1D simulation. The bottom panel of Fig.~\ref{fig:rsh_cluster} shows a comparison of the parameter $\Gamma$ measured in the 1D and 3D simulations (here, $\Gamma\equiv\dd\ln M_{\rm 200m}/\dd\ln a$; see also Eq.~\ref{eq:mar}). Both curves show a violent growth of the cluster at high redshift $z\gtrsim3$. But at lower redshift, the 1D curve becomes relatively smooth. Two prominent merger events appeared in the 3D simulation are absent in the 1D case.

\begin{figure}
\centering
\includegraphics[width=\linewidth]{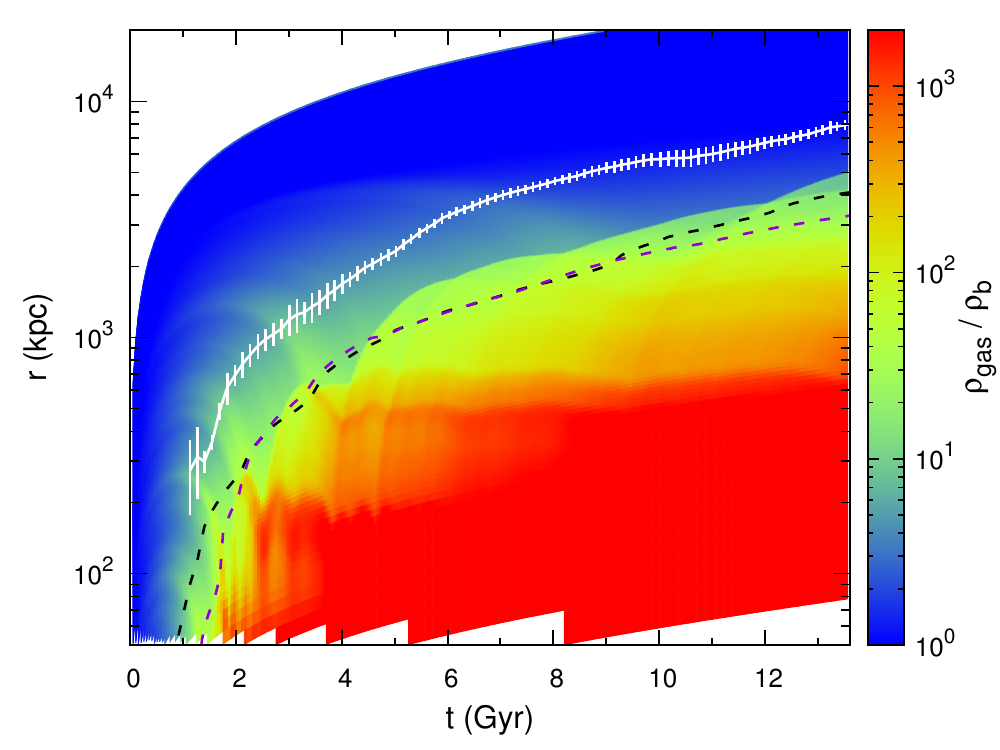}
\caption{Evolution of gas density profile in the 1D simulation (similar to those shown in Figs.~\ref{fig:1d_map}~and~\ref{fig:1d_map_appendix} but for the simulation with a more realistic initial condition). The dashed black and purple lines show the virial radii of the 3D and 1D clusters, respectively. As a comparison, the averaged shock radius of the 3D cluster along the non-filamentary directions is shown as the white line. Error bars represent the $1\sigma$ scatter. This figure shows that the 3D cluster has a much larger shock radius than that in the 1D cluster. Cluster mergers play a key role in this difference (see Section~\ref{sec:3d_simulations:comparison}).}
\label{fig:1d_cluster}
\end{figure}

Fig.~\ref{fig:1d_cluster} shows the evolution of the gas density profile in the 1D simulation. Even though only radial accretion is included, the growth of the gas halo is still not smooth. One can see a few high-density shells in the ICM, which are the angularly averaged infalling clumps. They compress the ICM and drive inner shocks inside the halo. As described in Section~\ref{sec:1d_model}, these inner shocks eventually encounter the accretion shock and expand the cluster shock-heated atmosphere. Nevertheless, this ``merger effect'' has been much weaker than that in the 3D situation. In other words, mergers and smooth accretion are distinguishable in this regard. As a reference, the black and purple lines in Fig.~\ref{fig:1d_cluster} show the cluster virial radius $R_{\rm 200m}$ in the 3D and 1D simulations, respectively. During $t=2-9\Gyr$, the virial radius of the 1D cluster shows a good match with that of the 3D calculation, but becomes $30\%$ smaller when $t>9\Gyr$, because of the absence of the two major mergers (see bottom panel in Fig.~\ref{fig:rsh_cluster}). The averaged shock radius, along the non-filamentary directions of the 3D cluster, is shown as the white line\footnote{Note that this curve is averaged over the entire cluster surface (excluding the filaments) but not only in the $x-y$ plane shown in Fig.~\ref{fig:3d_map}.}. The shock radius in the 1D model is found to be much smaller than this curve throughout the simulation. The $r_{\rm mas}/R_{\rm 200m}$ (or $r_{\rm acc}/R_{\rm 200m}$) is about $2.3$ and $1.3$ in our 3D and 1D clusters, respectively. The former value is generally consistent with that reported in \citet[][see also \citealt {Lau2015}]{Walker2019}; and the latter one agrees with that in the self-similar model \citep{Shi2016b}. The mismatch between the 1D and 3D clusters illustrates the importance of cluster mergers on the expansion of the boundary of the ICM through the shock collisions described in Section~\ref{sec:1d_model}, and also provides a robust way to distinguish the ordinary accretion shocks and \MAS{}s in the cosmological simulations and future observations. Furthermore, the DM splashback radius approximately aligns with the shock radius in our 1D simulation, i.e. $r_{\rm sp}\simeq r_{\rm acc}\simeq 1.3R_{\rm 200m}$\footnote{However, it is not a trivial task to measure the splashback radius for an individual 3D cluster because of the complicated merger and accretion configurations \citep{Mansfield2017}.}. It is interesting to note that both 3D cosmological simulations and the self-similar model show that $r_{\rm sp}\simeq 0.8-1.5R_{\rm 200m}$ depending on the mass accretion rate parameter $\Gamma$ ($\simeq0-5$; see \citealt{Lau2015,More2015,Mansfield2017} and also \citealt{Shi2016a}). The mergers and radial accretion play similar roles in changing $r_{\rm sp}$, which are quite different from that for the shock radii. This fact explains the misalignment between the shock and splashback radii in galaxy clusters.

\section{Conclusion and Discussion} \label{sec:conclusion}

In this work, we have explored the encounter of the merger and accretion shocks in galaxy clusters through the 1D and 3D cosmological simulations. As one of the key players, the runaway merger shocks usually exist in the diffuse gas between the high-density filaments \citep[see][]{Zhang2019b}. During the shock collisions, merger-accelerated accretion shocks (\MAS{}s) are formed and quickly propagate to larger cluster radii.  A notable signature of the shock collisions is that high-entropy shells are generated in between the \MAS{} fronts and the CDs (see Fig.~\ref{fig:1d_profiles}). These entropy structures further excite hydrodynamic instabilities and contribute to the non-thermal pressure in the cluster outskirts. Generally, the same type of MA-shocks should be present for a wide range of halo masses, provided that radiative cooling does not affect the gas and a hot atmosphere forms naturally. Basically, genuine accretion shocks should be present only during relatively quiescent periods of the halo evolution, while, long after each significant merger, the ``outer'' shocks would be located far outside the virial radius (between the filaments) and would be powered by the merger rather than by accretion.

The evolution of the \MAS{}s depends on the Mach number of the runaway merger shocks $\mathcal{M}_{\rm rs}$ and the cluster mass accretion rate parameter $\Gamma$. We found that, the \MAS{} fronts could reach very large cluster radii (i.e. $2-3$ times larger than those of the ordinary accretion shocks; see Fig.~\ref{fig:rsh_pred_scaled}) if $\Gamma\lesssim3$ and $\mathcal{M}_{\rm rs}\gtrsim1.5$. These conditions imply that \MAS{}s are not rare in galaxy clusters and could make strong impacts on the ICM. As the \MAS{} fronts represent the outer boundaries of the ICM, the cluster gas atmospheres are prominently expanded after the shock collision. The formation of \MAS{}s thus provides a natural explanation for the misalignment of the shock radii and the splashback radii found in the cosmological simulations \citep[e.g.][]{Lau2015}.

Since the \MAS{}s always reside beyond the cluster virial radius $R_{\rm 200m}$, it is a big challenge to detect them through their X-ray signals. The Sunyaev-Zel'dovich (SZ) effect, which linearly scales with the integrated electron pressure, provides a unique opportunity to probe the hot but low-density ICM in cluster outskirts \citep{Hurier2019}. However, we have to note that, in the cluster outer region, the electron-ion equilibrium timescale could be very long (e.g. a few $\Gyr$ or even longer; see \citealt{Avestruz2015}). The non-equilibrium electrons would blur the imprints of \MAS{}s on the cosmic microwave background (CMB). Nevertheless, given the nature of collisionless shocks, \MAS{}s are ideal targets for the next generations of the X-ray and SZ instruments (e.g. $AXIS/Lynx$; see \citealt{Mushotzky2019,Vikhlinin2019}) to explore the plasma physics of the ICM, e.g. magnetothermal instabilities, acceleration mechanisms of cosmic rays.

Furthermore, \MAS{}s are expected to play an important role in boosting radio emissivity of fossil relativistic electrons beyond the cluster virial radius \citep{Ensslin1998,Ensslin2001,Lyskova2019,Zhang2019b}. First, \MAS{}s pass through a very large volume of the intergalactic medium (IGM). The swept gas (also the fossil electrons) tends to accumulate behind the shock fronts (see top panel of Fig.~\ref{fig:1d_profiles}). Secondly, the low-efficiency issue of diffuse shock acceleration (DSA) in merger shocks is not a problem for the \MAS{}s any more \citep[see][and references therein]{Weeren2019}. The \MAS{}s' Mach number could reach a few tens to hundreds depending on the pre-heating process of the IGM.

\section*{Acknowledgments}
The  simulations  performed at the Leibniz-Rechenzentrum are under the project {\it pr86re}.
KD acknowledges support by the Deutsche Forschungsgemeinschaft (DFG, German Research Foundation) under Germany's Excellence Strategy - EXC-2094 - 390783311. EC acknowledges partial support by the Russian Science Foundation grant 19-12-00369. WF acknowledges support from the Smithsonian Institution and the High Resolution Camera program, part of the Chandra X-ray Observatory Center, which is operated by the Smithsonian Astrophysical Observatory for and on behalf of the National Aeronautics Space Administration under contract NAS8-03060. IZ is partially supported by a Clare Boothe Luce Professorship from the Henry Luce Foundation.

\appendix

\section{1D Cosmological Simulations} \label{sec:appendix:1d_simulation}

Our 1D cosmological simulation traces the evolution of both gas and DM in the self-gravitational field. A second-order piecewise-parabolic method (PPM) is applied to solve the Euler equations of the gas \citep{Fryxell2000} on a 1D spherical grid. The DM is however modelled as a series of collisionless shells, which are advanced in time through the leapfrog scheme \citep{Ricker2000}. In each time step, the DM shells are mapped to the grid to update the gravitational potential felt by the gas.

In all our simulations listed in Table~\ref{tab:simulation_params}, we used $200$ gas cells and $80000$ DM shells. To simultaneously reach a high spatial resolution for the cluster and suppress the boundary effect on the large-radius side, we arranged 160 cells uniformly to cover the innermost $10^3\,$comoving~kpc (ckpc), and the remaining 40 cells to cover $10^3-10^4{\rm\,ckpc}$ in the logarithmic scale\footnote{In Section~\ref{sec:3d_simulations:comparison}, we present a special 1D simulation, which is used to compare with the 3D cosmological simulation directly. In this simulation, we used $400$ gas cells and $80000$ DM shells. The size of the computational domain reaches $3\times10^{4}{\rm\,ckpc}$. The cosmological parameters adopted in this run are the same as those used in the 3D simulation (different from other 1D simulations presented in this paper).}. All cells contain the same number of DM shells in the initial condition. We set the initial gas and DM density profiles so that the cluster grows with a constant mass accretion rate $\Gamma$ (see Eq.~\ref{eq:mar}). Meanwhile, both DM and gas have zero initial velocity. All these simulations start from redshift $z=100$. We assumed a flat $\Lambda$CDM cosmology model with $\Omega_{\rm m0} = 0.30$, $\Omega_{\rm b0} = 0.05$, $\Omega_{\rm \Lambda0} = 0.70$, and $H_0 = 70\kms\Mpc^{-1}$ in the calculation. The results of the simulations S1 and S3 show good match with the self-similar solutions \citep{Fillmore1984}. We have also checked the convergence of our simulations in the spatial and mass resolutions. The run with the doubled numbers of gas cells and DM shells shows generally consistent results with those of the low-resolution run, but the excited discontinuities get sharper.

\begin{figure}
\centering
\includegraphics[width=\linewidth]{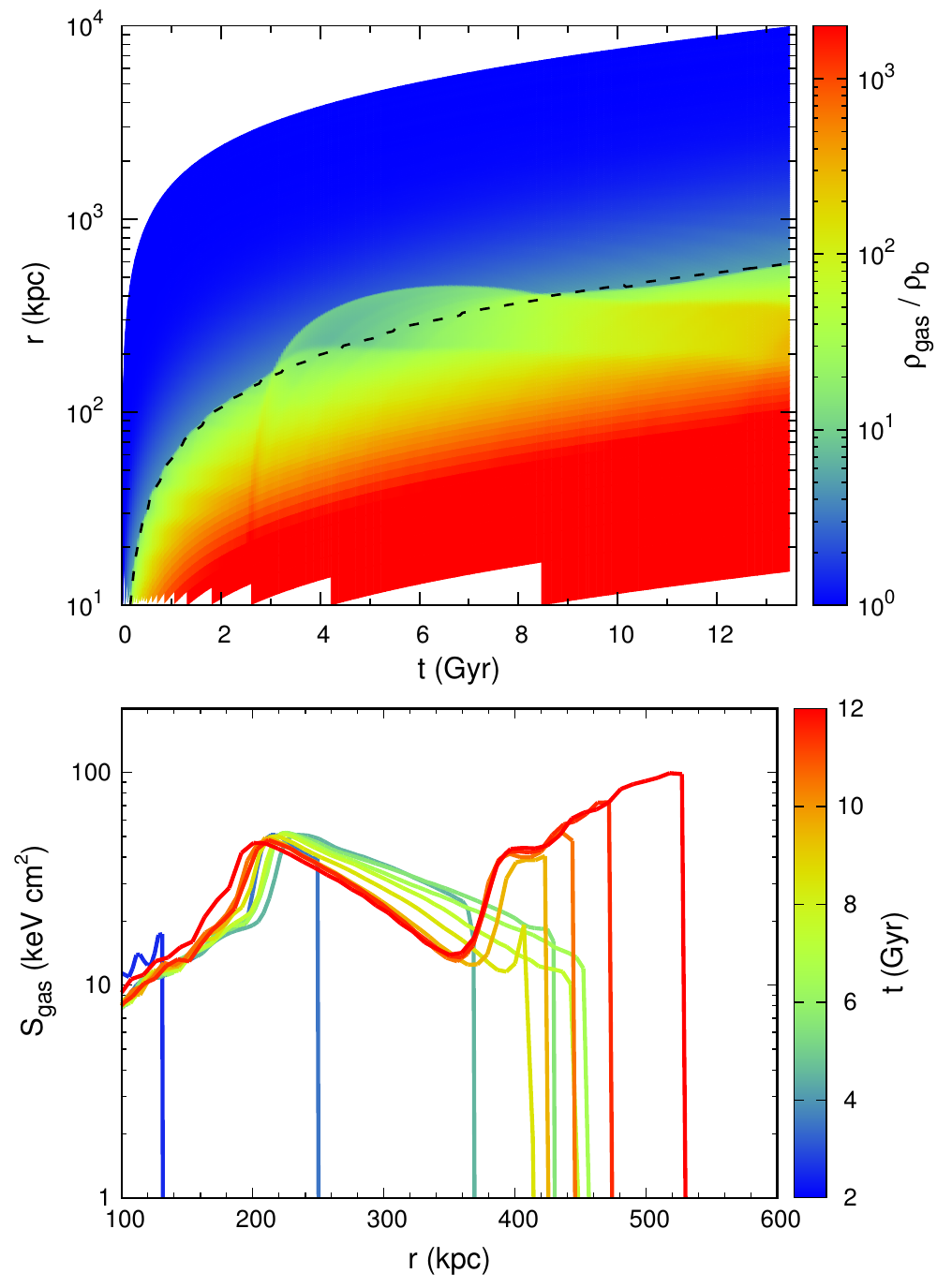}
\caption{\emph{Top panel:} Similar to Fig.~\ref{fig:1d_map} but for the gas density distribution from the simulation S1T2M15. \emph{Bottom panel:} Evolution of the corresponding gas entropy profile. This figure shows complementary results to those presented in Figs.~\ref{fig:1d_map}~and~\ref{fig:1d_profiles} (where the runaway shock has a larger Mach number). This figure shows that the \MAS{} has a shorter lifetime when the runaway shock is weaker (see Section~\ref{sec:1d_model:formation}). }
\label{fig:1d_map_appendix}
\end{figure}

\bsp	
\label{lastpage}
\end{document}